\documentclass[12pt]{article}

\usepackage[super,comma,numbers,sort&compress]{natbib}
\usepackage{graphicx}
\usepackage{authblk}
\usepackage[indention=0pt]{subfig}
\usepackage{amsmath}
\usepackage{amssymb}
\usepackage{xcolor}
\usepackage{times}
\usepackage{braket}
\usepackage[colorlinks=false, pdfborder={0 0 0}]{hyperref}
\usepackage{placeins}
\title{Collective photon emission patterns from two atoms in free space}
\author[*,1,2]{S. Richter}
\author[3]{S. Wolf}
\author[1,2]{J. von Zanthier}
\author[3]{F. Schmidt-Kaler}
\affil[1]{Institut f\"ur Optik, Information und Photonik, Friedrich-Alexander Universit\"at Erlangen-N\"urnberg, Staudtstr. 1, 91058 Erlangen, Germany}
\affil[2]{Erlangen Graduate School in Advanced Optical Technologies (SAOT), Friedrich-Alexander Universit\"at Erlangen-N\"urnberg, Paul-Gordan-Str. 6, 91052 Erlangen, Germany}
\affil[3]{QUANTUM, Institut f\"ur Physik, Johannes Gutenberg-Universit\"at Mainz, Staudingerweg 7, 55128 Mainz, Germany}
\affil[*]{To whom correspondence should be addressed. E-mail: stefan.michael.richter@fau.de.}

\date{}

\begin{document}

\maketitle

{\bf
Modification of spontaneous decay in space and time is a central topic of quantum physics.
It has been predominantly investigated in the context of cavity quantum electrodynamics (QED) \cite{bates1985advances}, gaining new interest recently in the domain of nano-optics \cite{zoller2017,chang2018colloquium}.
Beyond cavity-QED, spontaneous emission may be modified also in free space due to correlations among the photon emitters,
a phenomenon known as super- and subradiance \cite{dicke1954coherence,agarwal1974quantum,gross1982superradiance}.
Correlations may stem either from direct interactions between the particles \cite{devoe1996observation,blatt2001,monroe1995demonstration,monz201114,tan2015multi,ballance2015hybrid,Bloch2020subradiant,Browaeys2021}, from long-range exchange of photons \cite{Walraff2013,Lukin2016,Lukin2018,Waks2018,Gammon2019,Rauschenbeutel2020},
or by measuring single photons in a common mode \cite{cabrillo1999creation,skornia2001nonclassical,Plenio2003,Irvine2003,moehring2007entanglement,hofmann2012heralded,slodivcka2013atom,bernien2013heralded,mcconnell2015entanglement,delteil2016generation,araneda2018interference,wolf2020light}.
Yet, the genuine spatial spontaneous emission pattern of an atomic ensemble in an entangled quantum state has not been observed so far, due to the lack of ultra-fast cameras with high spatial resolution suited for recording single photons from single atoms.
Preparing two trapped ions in free space in entangled Dicke states via photon detection, we study the resulting collective spontaneous emission patterns. Depending on the symmetry of the Dicke states, associated with the direction of detection of the first state-determining photon, we observe fundamentally different emission patterns for the subsequently scattered photon,
including super- and subradiance.
Our results demonstrate that the detection of a single photon can profoundly modify the collective emission of an atomic array, here represented by its most elementary building block of two atoms in free space.
}
\vspace{1cm}

It is well established that spontaneous emission of radiation is not an immutable property of an atom, but can be modified, e.g., by particular boundary conditions of the electromagnetic field as in cavity-QED \cite{bates1985advances}. Another option to change the spatio-temporal characteristics of spontaneous decay exploits particle-particle correlations in an atomic ensemble leading to collective light emission known under the name of super- and subradiance \cite{dicke1954coherence,agarwal1974quantum,gross1982superradiance}. Correlations among the emitters can be induced by direct interactions if the interatomic separations are smaller than the transition wavelength, e.g., via exchange of virtual photons \cite{devoe1996observation,blatt2001,Kimble2015,Ruostekoski2016,Adams2016,Yelin2017,Kimble2017,Volz2017,Browaeys2021}. Such cooperative response has been observed recently in the directional reflection of light by a two-dimensional array of atoms trapped in an optical lattice, including spectral narrowing associated with subradiance \cite{Bloch2020subradiant}. For larger interatomic separations, correlations among the particles can be generated via the dynamical exchange of real photons mediated by nano-optical devices \cite{zoller2017,chang2018colloquium}, e.g., using photonic wave-guides \cite{Walraff2013,Lukin2016,Lukin2018,Waks2018,Gammon2019,Rauschenbeutel2020}.

An alternative method for generating correlations among particles relies on the measurement of photons
excluding which-path-information \cite{cabrillo1999creation,skornia2001nonclassical,Plenio2003,Irvine2003,thiel2007generation,moehring2007entanglement,bastin2009operational,hofmann2012heralded,slodivcka2013atom,bernien2013heralded,oppel2014directional,Wiegner2015,mcconnell2015entanglement,delteil2016generation,araneda2018interference,wolf2020light}. This  can be realized with indistinguishable photons scattered by the emitters into a common mode, i.e., mixing the photons in a beam splitter \cite{Plenio2003,moehring2007entanglement,hofmann2012heralded,slodivcka2013atom,bernien2013heralded,delteil2016generation,araneda2018interference} or recording the photons in the far field \cite{cabrillo1999creation,skornia2001nonclassical,Irvine2003,thiel2007generation,bastin2009operational,oppel2014directional,Wiegner2015,mcconnell2015entanglement,wolf2020light}.
Based on this scheme, the generation of the fully symmetric as well as the fully anti-symmetric Dicke state has been suggested for two atoms initially prepared in the excited state, depending on the direction of observation of the first scattered photon \cite{skornia2001nonclassical} (see Fig. 1a and method section).
Subject to which state is produced, superradiant and subradiant spatial emission patterns are predicted for the subsequently radiated photon \cite{wiegner2011quantum}.
Employing more atoms, entangled Dicke states with higher number of excitations have been proposed using the same technique \cite{thiel2007generation,bastin2009operational}, displaying more complex emission patterns \cite{oppel2014directional,Wiegner2015,wiegner2011quantum}.
So far, entanglement via projective measurements of photons has been implemented with ions \cite{moehring2007entanglement,slodivcka2013atom,araneda2018interference,wolf2020light}, neutral atoms \cite{hofmann2012heralded,mcconnell2015entanglement}, NV-centers \cite{bernien2013heralded}, and quantum dots \cite{delteil2016generation,Atature2017}. Yet, in all of these cases single spatial modes have been picked out for the measurement of the subsequently scattered photons, inhibiting  the observation of a genuine spatial emission pattern of the spontaneous radiation.

Here we report on experimentally projecting two distant atoms onto entangled Dicke states via  photon detection and observing
the collective spontaneous  emission pattern of the subsequently emitted photon.
Recording two subsequent photons from the atomic ensemble
amounts to measuring the second-order photon correlation function. Assuming two laser-driven two-level atoms and coincident detection, the normalized second-order photon correlation function is given by \cite{skornia2001nonclassical}
\begin{equation}
g^{(2)}(\boldsymbol{r}_1,\boldsymbol{r}_2, \tau = 0) = \frac{(1+s)^2 \, \cos^2\left(\left(\delta(\boldsymbol{r}_1)-\delta(\boldsymbol{r}_2)\right)/2\right)}{\left(1+s+\cos\delta(\boldsymbol{r}_1)\right)\left(1+s+\cos\delta(\boldsymbol{r}_2)\right)}.
\label{eq:g2Theo}
\end{equation}
Here, $s$ denotes the saturation parameter of the atomic transition $\ket{g}\leftrightarrow\ket{e}$
and $\delta(\boldsymbol{r})=(\boldsymbol{k}_L-k_L\hat{\boldsymbol{r}})\cdot\boldsymbol{d}$
the optical phase difference of a photon recorded at position $\boldsymbol{r}$ if scattered by atom $1$ or atom $2$, respectively, with $\boldsymbol{k}_L$ the laser wave vector and $\boldsymbol{d}$ the distance vector of the two atoms (see method section for details).

The two-photon correlation function $g^{(2)}(\boldsymbol{r}_1,\boldsymbol{r}_2, \tau = 0)$ displays a modulation as a function of $\delta(\boldsymbol{r}_2)$ if the second detector at position $\boldsymbol{r}_2$ is moved along an axis parallel to the ion-distance vector $\boldsymbol{d}$, with an initial phase depending on $\delta(\boldsymbol{r}_1)$, i.e., the position of the first photon detection event (see Eq.~(\ref{eq:g2Theo})). In particular, $g^{(2)}(\boldsymbol{r}_1,\boldsymbol{r}_2, \tau = 0)$ exhibits a maximal detection probability in the direction $\delta(\boldsymbol{r}_2)=0$ for the second photon if the first photon has been also recorded in the direction $\delta(\boldsymbol{r}_1)=0$, corresponding
to spatial superradiance (see Fig. 1 and method section) \cite{oppel2014directional,Wiegner2015,wiegner2011quantum}. By contrast, $g^{(2)}(\boldsymbol{r}_1,\boldsymbol{r}_2, \tau = 0)$ features a minimum emission probability for the second photon along $\delta(\boldsymbol{r}_2)=0$ if the first photon has been recorded at a position such that $\delta(\boldsymbol{r}_1)=\pi$, corresponding to spatial subradiance \cite{wiegner2011quantum}.

To investigate these collective emission properties of two atoms in free space,
we store two $^{40}$Ca$^+$ ions in a segmented linear Paul trap with trap frequencies $\omega_\text{ax, Rad1, Rad2}/2\pi=(0.76, 1.275,\\ 1.568)\,$MHz for the axial and the two radial modes, respectively. The trapped ions are continuously excited and laser-cooled on the $S_{1/2} - P_{1/2}$ transition by a laser near $397\,$nm (see Fig. 1(b) and 1(c)). A quantization axis is defined by a magnetic field $B=0.62\,$mT oriented along the $z$-direction, i.e., vertical to the plane defined by the two vectors $\boldsymbol{r}_1$ and $\boldsymbol{r}_2$ and coinciding with the polarization of the exciting laser. A $f/1.6$ lens system collects $2.5\% $ of the photons scattered by the two ions on the $S_{1/2} - P_{1/2}$ transition.
It directs the light onto a polarization filter, equally oriented along the $z$-direction, ensuring indistinguishably of the scattered photons. The photons are fed thereafter into a Hanbury-Brown and Twiss (HBT) setup consisting of a non-polarizing $50/50$ beam splitter and  two synchronized ultrafast microchannel plate (MCP) detectors with $1000\times1000$ virtual spatial bins and a timing resolution of $50\,$ps, combining very high spatial and temporal resolution. With a MCP dead time of $600$~ns, the setup requires two MCP detectors to determine the second order photon-correlation function $g^{(2)}(\boldsymbol{r}_1,\boldsymbol{r}_2, \tau)$ also within the coherence time of the ions. Here, $\boldsymbol{r}_1$ and $\boldsymbol{r}_2$ denote the position of the first and the second photon detection event on the first and the second MCP camera, respectively, and $\tau$ the difference in arrival time of the two photons.

Measuring with this setup the photon stream on the two MCP cameras for about $205$ consecutive hours, $g^{(2)}(\boldsymbol{r}_1,\boldsymbol{r}_2, \tau)$ is computed by reducing the two-dimensional $1000\times1000$ spatial data set of each frame of each MCP camera to $1 \times 96$ bins. This increases the statistics per bin taking advantage of the fact that the relevant spatial modulation of $g^{(2)}(\boldsymbol{r}_1,\boldsymbol{r}_2, \tau)$ occurs in one dimension only, i.e., parallel to the ion-distance vector $\boldsymbol{d}$ (see Fig. 1c).
We further choose time bins of $2.5$ ns, significantly longer than the time resolution of the MCP cameras but shorter than the lifetime of the excited state of $^{40}$Ca$^+$ of $\tau_{P_{1/2}} = 6.9$ns \cite{Poschinger2015}. The total of the collected two-photon correlation data is thus stored in a $96 \times 96 \times 38$ bin data structure, encoding the position of the first (second) photon detection event at $\boldsymbol{r}_1$ (at $\boldsymbol{r}_2$) as well as the photon arrival time difference $\tau$, with each entry of the data structure filled on average by about $20$ events.



The normalized photon correlation function $g^{(2)}(\boldsymbol{r}_1,\boldsymbol{r}_2, \tau)$ is displayed in Fig.~2(a) as a function of $\delta(\mathbf{r}_2)$ and the photon arrival time difference $\tau$
for four different positions of the first photon detection event, i.e., for $\delta(\mathbf{r}_1)= 0.34 \pi, 0.73\pi, 1.05 \pi$, and $1.38 \pi$.
Cuts of the histograms at $\tau = 0$ are shown in Fig.~2(b), binned to two periods $\delta(\mathbf{r}_2) \in [-2 \pi, + 2 \pi]$.
As can be seen from the plots, Figs.~2(a) and 2(b) display indeed completely distinct spatial patterns for the emission probability of the second spontaneously scattered photon as a function of $\delta(\mathbf{r}_1)$, i.e., the position of the first photon detection event.
In particular, we recognize a pattern with maximum emission probability of spontaneously scattered radiation in the direction $\delta(\mathbf{r}_2) \approx 0$ in the case that the first photon has been recorded at $\delta(\mathbf{r}_1)= 0.34 \, \pi$ (blue dots in Fig.~2(b)(i)). For this setting the two-ion system has been projected (approximately) into the symmetric Dicke state  $\ket{s}$. This is the regime of spatial superradiance
\cite{wiegner2011quantum,oppel2014directional,Wiegner2015}.
By contrast, we observe a pattern with minimum emission probability of spontaneously scattered radiation in the direction $\delta(\mathbf{r}_2)\approx 0$ if the first photon  has been recorded at $\delta(\mathbf{r}_1)=  1.05 \, \pi$ (red dots in Fig.~2(b)(i)). In this case, the two-ion system has been projected into the anti-symmetric Dicke state $\ket{a}$, corresponding to the regime of spatial subradiance (see Fig. 1(a) and method section) \cite{wiegner2011quantum}.
Note that in Fig.~2(b)(i) the slight deviations of $\delta(\mathbf{r}_1)$ from the more transparent values $\delta(\mathbf{r}_1)= 0$ and $\delta(\mathbf{r}_1)= \pi$ result from the spatial binning process outlined above. However, this does not impair the general observation that the different emission patterns for the second spontaneously emitted photon depend uniquely on the position of the first photon detection event, i.e., on $\delta(\mathbf{r}_1)$, determining the state onto which the two-ion system is projected on \cite{wiegner2011quantum}.
The different spatial patterns in Fig.~2(b) thus clearly demonstrate that the detection of a single photon profoundly modifies the spatial emission characteristics of the subsequently scattered photon, i.e., the collective spontaneous photon emission of the atomic array, here represented by its most elementary building block of two atoms in free space.

In addition to super- and subradiance, the photon auto-correlation function of the two-ion system
displays anti-bunching $g^{(2)}(\boldsymbol{r}_1,\boldsymbol{r}_1, \tau) < 1$ for $\delta(\mathbf{r}_1)= \delta(\mathbf{r}_2)= 0.34 \, \pi$  and bunching $g^{(2)}(\boldsymbol{r}_1,\boldsymbol{r}_1, \tau) > 1$ for $\delta(\mathbf{r}_1)= \delta(\mathbf{r}_2)= 1.05 \, \pi$ (see Fig.~2(c)).
This results from the fact that laser light couples only the symmetric states $\ket{g,g}$, $\ket{s}$ and $\ket{e,e}$ of the two-ion system, not the anti-symmetric state $\ket{a}$. Thus, for low saturation of the continuously exciting laser, essentially only the two-ion ground state $\ket{g,g}$ is occupied, whereas the state $\ket{s}$ is much less populated and the state$\ket{e,e}$ hardly at all. For $\delta(\mathbf{r}_1)= \delta(\mathbf{r}_2) \approx 0$, i.e., when the measured photons may only stem from the symmetric decay channels $\ket{e,e}\rightarrow\ket{s}$ or $\ket{s}\rightarrow\ket{g,g}$ (see Fig.~1(a) and method section), a recorded photon will in most cases originate from the transition s$\ket{s}\rightarrow\ket{g,g}$. Hence, before measuring the next photon, the system has to be reexcited to the state $\ket{s}$, resulting in anti-bunching of the photon stream \cite{wolf2020light}.
By contrast, for $\delta(\mathbf{r}_1)= \delta(\mathbf{r}_2) \approx \pi$,
the auto-correlation $g^{(2)}(\boldsymbol{r}_1,\boldsymbol{r}_1, \tau)$ measures only photons decaying via the anti-symmetric decay channel  $\ket{e,e}\rightarrow\ket{a}\rightarrow\ket{g,g}$ (see Fig.~1(a) and method section) corresponding to a consecutive photon emission, resulting in bunching of the scattered photons \cite{wolf2020light}.
In this way, the maximum of superradiance, i.e., a second-order correlation function $g^{(2)}(\boldsymbol{r}_1,\boldsymbol{r}_1, \tau =0)$ observed at $\delta(\mathbf{r}_2) = \delta(\mathbf{r}_1) \approx 0$, appears in combination with photon anti-bunching,
whereas the maximum of spatial subradiance, i.e., the maximum of $g^{(2)}(\boldsymbol{r}_1,\boldsymbol{r}_1, \tau =0)$ appearing at $\delta(\mathbf{r}_2)= \delta(\mathbf{r}_1) \approx \pi$, goes along with photon bunching.

In conclusion, we investigated the collective spontaneous emission behavior of a correlated atomic array in free space, represented by its most elementary building block of two trapped ions. This fundamental and conceptually simple setup leads to a great wealth of spatio-temporal emission phenomena characterized by measuring the second order photon correlation function $g^{(2)}(\boldsymbol{r}_1,\boldsymbol{r}_2, \tau)$. The direction of detection of the first photon, determining uniquely the state the two-ion system is projected on, fully governs the emission pattern of the second spontaneously scattered photon. This spatial emission pattern yields on the one hand structural information about the sources \cite{richter2021imaging}, on the other hand it allows for a unique identification of the corresponding state. Yet beyond, the direction of detection of the first photon also controls the second photon’s temporal detection characteristics, undergoing a continuous crossover from photon anti-bunching to bunching. We measured these genuine features of collective spontaneous decay and fully model the observed behavior by use of the Dicke basis, taking independently derived experimental parameters into account. The clean and versatile model system thus displays how quantum cooperativity is established via projective measurements. In the future, we will explore the collective spontaneous emission characteristics for larger ion crystals, eventually extending to two-dimensional arrays, aiming for projective preparation and imaging of long-lived entangled states.

\paragraph{Data availability}
Source data for the plots shown is provided under https://dx.doi.org/10.22000/560.
The experimental data that support the findings of this study are available
from the corresponding author upon reasonable request.


\paragraph{Acknowledgments.} S.R. and J.v.Z. acknowledge support from the Graduate School of Advanced Optical Technologies (SAOT) and the International Max-Planck Research School, Physics of Light, Erlangen. We thank Photonscore GmbH, Brenneckestr. 20, 39118 Magdeburg, Germany
for providing the synchronized MCP camera system (https://photonscore.de) and
Andr\'e Weber for the initial calibration and characterization of the MPC
systems. J.v.Z. thanks Ralf Palmisano for making contact to Photonscore GmbH. This research is funded by the Deutsche Forschungsgemeinschaft (DFG, German Research Foundation) within the TRR 306 QuCoLiMa (``Quantum Cooperativity of Light and Matter'') -- Project-ID 429529648.

\paragraph{Author contribution}

J.v.Z. and F.S.-K. conceived the concept of the experiment. The measurements were performed by S.R. and S.W. All Authors worked on the analysis and interpretation of the data and contributed to the manuscript.

\bibliographystyle{naturemag}
\bibliography{litnat}

\newpage

\begin{figure}[ht!]
\begin{center}
\includegraphics[width=100mm]{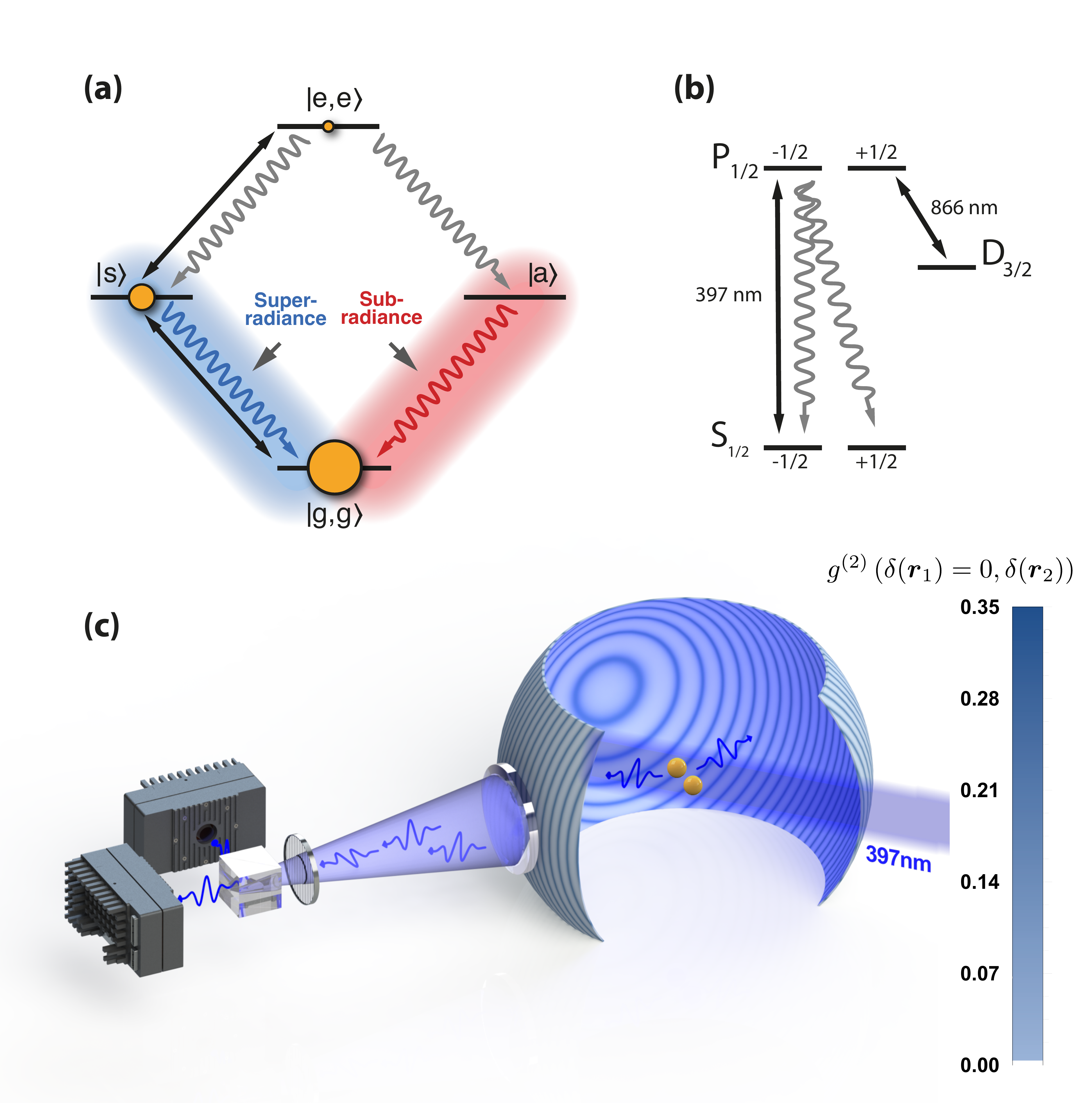}
\end{center}
\caption{
(a) Level  scheme of the two-ion system using the Dicke-basis; starting from the
  fully excited state $\ket{e,e}$ and measuring a spontaneously emitted photon
  in the far field at $\boldsymbol{r}_1$ such that $\delta(\boldsymbol{r}_1)=n
  \cdot 2 \pi$, $n \in N$, projects the two ion-system into the symmetric Dicke
  state $\ket{s}$ (see method section); from $\ket{s}$, the two-ion system emits
  a second photon with a spatial distribution corresponding to superradiance
  (blue transition, see blue dots in Fig 2(b)(i)); by contrast, measuring a
  photon at $\boldsymbol{r}_1$ such that $\delta(\boldsymbol{r}_1)=(2n+1) \cdot
  \pi$, $n \in N$, the two ion-system is projected into the anti-symmetric Dicke
  state $\ket{a}$; from $\ket{a}$, the two-ion system emits a second photon with
  a spatial distribution corresponding to subradiance (red transition, see red dots in Fig 2(b)(i)); laser light (black double-headed arrows) couples only the symmetric Dicke states $\ket{g,g}$, $\ket{s}$, $\ket{e,e}$ of the two-ion system; hence, for low laser saturation, essentially only the two-ion ground state $\ket{g,g}$ is populated, whereas the states $\ket{s}$ and $\ket{e,e}$ are hardly populated at all (see yellow circles).
(b) Relevant level scheme of $^{40}$Ca$^+$: a $\pi$-polarized laser drives the $S_{1/2} - P_{1/2}$ transition near $397\,$nm (black double-headed arrows); the $P_{1/2}$ state with a life time $\tau_{P_{1/2}} = 6.9$ns decays back into the ground state $S_{1/2}$ with $93.6\,\%$ probability and with $6.4\,\%$ probability into the metastable $D_{3/2}$ state; a repumping laser near $866\,$nm repopulates the ions from the $D_{3/2}$ state to the $P_{1/2}$ state.
(c) Scheme of the experimental setup: two $^{40}$Ca$^+$ ions are stored in a linear Paul trap (not shown) and are continuously excited on the $S_{1/2} - P_{1/2}$ transition by a laser near $397\,$nm. $2.46\% $ of the  spontaneously scattered light is collected by a lens and, after passing a polarization filter to ensure indistinguishably of the scattered photons, is fed into a Hanbury-Brown and Twiss setup to measure $g^{(2)}(\boldsymbol{r}_1,\boldsymbol{r}_2, \tau)$; the Hanbury-Brown and Twiss setup consists of two synchronized microchannel plate detectors
combining high spatial and temporal resolution; the sphere around the two ions indicates the values for $g^{(2)}(\boldsymbol{r}_1,\boldsymbol{r}_2, \tau=0)$ measured for the case $\delta(\mathbf{r}_1)= 0.34 \pi$.
}
\label{fig:Kugel}
\end{figure}

\begin{figure}[ht!]
\centering
\includegraphics[width=150mm]{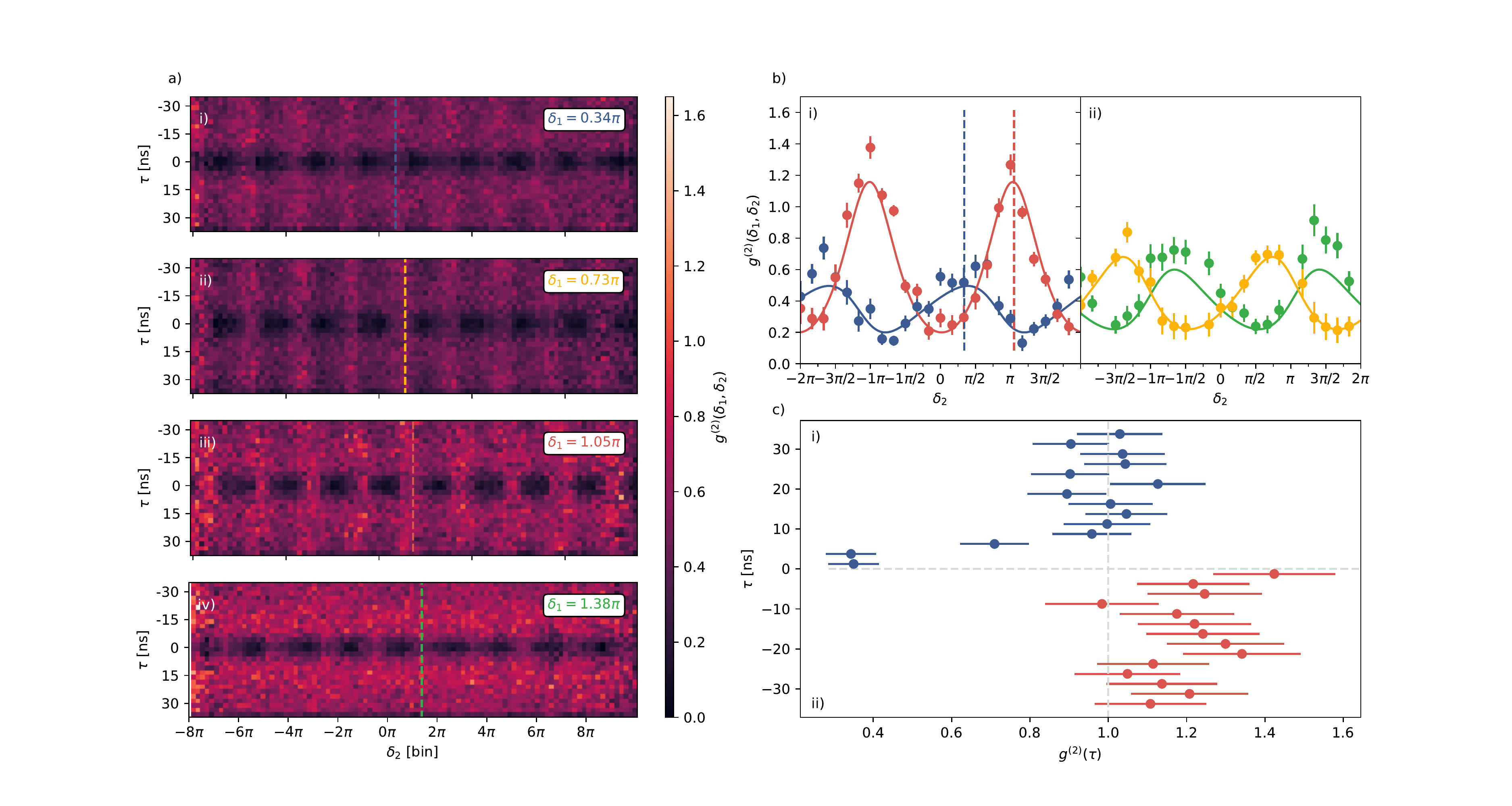}
\caption{
(a) Color-coded histograms of the spatio-temporal photon correlation function $g^{(2)}(\boldsymbol{r}_1,\boldsymbol{r}_2, \tau)$ as a function of $\delta(\mathbf{r}_2)$ and photon arrival time difference $\tau$ measured for four different directions of the first photon detection event (indicated as doted lines):  $\delta(\mathbf{r}_1)= 0.34 \pi$ (i), $\delta(\mathbf{r}_1)=0.73\pi$ (ii), $\delta(\mathbf{r}_1)=1.05 \pi$ (iii), $\delta(\mathbf{r}_1)=1.38 \pi$ (iv).
(b) Cuts of the histograms of Fig.~2(a) at $\tau = 0$, binned to $\delta(\mathbf{r}_2) \in [-2 \pi, + 2 \pi]$: (i) approximate spatial superradiance (blue dots) and subradicance (red dots), i.e., maximum (minimum) emission probability for the second spontaneously scattered photon emitted in the direction $\delta(\mathbf{r}_2) \approx 0$ in the case that $\delta(\mathbf{r}_1)= 0.34 \, \pi$ ($\delta(\mathbf{r}_1)=  1.05 \, \pi$);
(ii) intermediate regimes between super- and subradiance for the second spontaneously scattered photon in the case that $\delta(\mathbf{r}_1)=0.73\pi$ (yellow dots) and $\delta(\mathbf{r}_1)=1.38 \pi$ (green dots); error bars correspond in all cases to one statistical standard deviation; blue, red, yellow, and green solid lines are derived from Eq.~(3), taking into account the independently measured Debye-Waller factors $f_{DWF}^{\boldsymbol{r}_1} \approx f_{DWF}^{\boldsymbol{r}_2} = 0.51$, the saturation parameter $s = 0.65$, as well as an additional offset $\Delta \approx 0.2$ (see method section).
(c) Auto-correlation function $g^{(2)}(\tau)$ obtained from Fig.~2(a) for (i) $\delta(\boldsymbol{r_1}) = \delta(\boldsymbol{r_2}) = 0.34 \pi$ (indicated as blue vertical dotted line in Fig.~2(b)(i)) displaying photon anti-bunching, and (ii) $\delta(\boldsymbol{r_1}) = \delta(\boldsymbol{r_2}) = 1.05 \pi$ (indicated as red vertical dotted line in Fig.~2(b)(i)) displaying photon bunching; error bars correspond in both cases to one statistical standard deviation.}
\label{fig:rebinned}
\end{figure}

\FloatBarrier
\newpage

\vspace{1cm}

\textbf{Methods}

\vspace{0.5cm}

\textbf{Preparing Dicke states via photon detection}

\vspace{0.5cm}

A pair of identical atoms with separation $d \gg \lambda$, as is the case for the considered two-ion system, is best described using the Dicke basis, employing the symmetric and anti-symmetric combination of the ground states $\ket{g}_i$ and excited states $\ket{e}_i$ of the two ions $i = 1,2$, respectively (see Fig. 1a)
\begin{equation*}
\begin{aligned}
&\ket{e,e},\\
&\ket{s}=\frac{1}{\sqrt{2}}\left(\ket{e,g}+\ket{g,e}\right),\\
&\ket{g,g};\\
&\ket{a}=\frac{1}{\sqrt{2}}\left(\ket{e,g}-\ket{g,e}\right).
\end{aligned}
\end{equation*}
Starting from the fully excited state $\ket{e,e}$, after detection of a spontaneously emitted photon in the far field at position $\boldsymbol{r}$, the two-ion system is projected into the general state $\psi (\boldsymbol{r}) = \alpha \ket{s} + \beta \ket{a}$ with one excitation, with $|\alpha|^2 + |\beta|^2 =1$, where the prefactors $\alpha$ and $\beta$ depend on the position $\boldsymbol{r}$.
This can be seen
when considering the photon detection operator \cite{skornia2001nonclassical}
\begin{equation}
\hat{D}(\boldsymbol{r})=\hat{\sigma}_1+e^{i\delta (\boldsymbol{r})}\hat{\sigma}_2,
\label{eq:DecayOperator}
\end{equation}
expressing by use of the individual atomic lowering operators $\hat{\sigma}_1=\ket{g}_1\bra{e}$ and $\hat{\sigma}_2=\ket{g}_2\bra{e}$
the two possibilities that either the first or the second atom emits a photon via spontaneous decay. Since the photon detection occurs in the far field, the detection cannot distinguish whether the photon has been emitted by atom $1$ or atom $2$, hence both options have to be considered as in Eq.~(\ref{eq:DecayOperator}). Moreover, the operator $\hat{D}(\boldsymbol{r})$ takes into account the optical phase difference $\delta(\boldsymbol{r})$ between these two possibilities, expressing the difference in phase of a photon recorded at position $\boldsymbol{r}$ when scattered by atom $1$ or atom $2$, respectively.

If for both atoms initially in the state $\ket{e,e}$ the detector recording the first emitted photon is placed at a position such that $\delta(\boldsymbol{r}_1)$ is an even integer of $\pi$,
the detection operator $\hat{D}(\boldsymbol{r})$ is symmetric and connects the initial state $\ket{e,e}$ exclusively with the symmetric Dicke state $\ket{s}$, going along with $|\alpha|^2 = 1$
(see Fig. 1a). By contrast, if the detector is located at a position such that $\delta(\boldsymbol{r}_1)$ is an odd integer of $\pi$, the detection operator $\hat{D}(\boldsymbol{r})$  is antisymmetric and connects the initial state $\ket{e,e}$ only with the antisymmetric Dicke state $\ket{a}$, going along with $|\beta|^2 = 1$.
Moving the detector recording the first photon in space from $\delta(\boldsymbol{r}_1)= 0$ to $\delta(\boldsymbol{r}_1)= \pi$ will thus continuously modify the state $\psi (\boldsymbol{r})$ from $\ket{s}$ to $\ket{a}$.
Correspondingly, the spatial distribution of the second spontaneously emitted photon recorded by a detector at $\boldsymbol{r}_2$ will gradually change from superradiant to subradiant, while otherwise all parameters of the source are kept unchanged.
\\

\textbf{Computing the normalized second order correlation function from the raw data}\\
\\
After recording a two-dimensional $1000\times1000$ spatial data set for each frame of each MCP camera, we reduce the data of each frame to $1 \times 96$ bins and time bin the photon arrival time differences to $2.5$ ns. In this way, the total of the collected two-photon correlation data is stored in a $96 \times 96 \times 38$ bin data structure, encoding the position of the first (second) photon detection event at $\boldsymbol{r}_1$ ($\boldsymbol{r}_2$) as well as the photon arrival times difference $\tau$.
Recording photons on the two MCP cameras consecutively for about $205$ h, each entry of the data structure is filled on average with about $20$ events.

Recording in parallel also the intensities $G^{(1)}\left(\mathbf{r}_i\right)$ at each MCP camera $i$, $i = 1,2$, stored equally in a $1 \times 96$ bin data frame, we are able to compute the
normalized second order photon correlation function. For coincident detection it is defined as
$$
g^{(2)}\left( \mathbf{r}_1,
  \mathbf{r}_2, \tau=0 \right) = G^{(2)}\left( \mathbf{r}_1,
  \mathbf{r}_2, \tau=0 \right)/(G^{(1)}\left(\mathbf{r}_1\right) \cdot
  G^{(1)}\left(\mathbf{r}_2\right)).
$$
Normalization of the entire data set for arbitrary $\tau$ is finally accomplished
making use of the fact that $\lim_{\tau \to \infty} g^{(2)}\left( \boldsymbol{r}_1,\boldsymbol{r}_1, \tau \right) = 1$.\\
\\

\textbf{Comparing the measured normalized second order correlation function to theory}\\
\\
In order to model the experimentally measured data, Eq.~\eqref{eq:g2Theo} needs to be expanded to include the Debye-Waller factor $f_{DWF}^{\boldsymbol{r}}$ taking into account the residual motion of the trapped ions as well as the momentum transfer between the incoming and scattered photons and the ions. Allowing for an additional offset $\Delta \approx 0.2$ resulting from dark counts, repumping of the ions from the metastable $D_{3/2}$ state by an additional laser, and convolutions due to temporal and spatial binning, Eq.~\eqref{eq:g2Theo} becomes
\begin{equation}
g^{(2)}(\boldsymbol{r}_1,\boldsymbol{r}_2, \tau = 0) = \frac{(1+s)^2 \,
  \cos^2\left(\left(\delta(\boldsymbol{r}_1)-\delta(\boldsymbol{r}_2)\right)/2\right)f_{DWF}^{\boldsymbol{r}_1}f_{DWF}^{\boldsymbol{r}_2}}{\left(1+s+\cos\delta(\boldsymbol{r}_1)
  f_{DWF}^{\boldsymbol{r}_1} \right)\left(1+s+\cos\delta(\boldsymbol{r}_2)
  f_{DWF}^{\boldsymbol{r}_2} \right)} + \Delta,
\end{equation}
where $f_{DWF}^{\boldsymbol{r}_i}$ is given by \cite{PhysRevA.57.4176}
\begin{equation}
f_{DWF}^{\boldsymbol{r}_i} = \text{exp} \left[ -\frac{\hbar q_x^2(\boldsymbol{r}_i)}{m \omega_{r1}} \left( \langle N_{r1} \rangle + \frac{1}{2} \right) -\frac{\hbar q_y^2(\boldsymbol{r}_i)}{m \omega_{r2}} \left( \langle N_{r2} \rangle + \frac{1}{2} \right) -\frac{\hbar q_z^2(\boldsymbol{r}_i)}{m \omega_{b}} \left( \langle N_b \rangle + \frac{1}{2} \right) \right]
.\end{equation}
Here $q_{\alpha}(\boldsymbol{r}_i)$, $i = 1, 2$, describes the projection of the momentum transfer
vector $\boldsymbol{q}(\boldsymbol{r}_i) = \boldsymbol{k}_L - k_L
\hat{\boldsymbol{r}}_i$ onto the basis axis $\alpha = x, y, z$ of the underlying vibrational modes $\omega_{\kappa}$ occupied by $\left\langle N_{\kappa} \right\rangle$ phonons, where $\omega_{\kappa}$ are the frequencies of the relevant vibrational modes $\kappa=r1, r2, b$.
For the investigated two-ion system, only three contrast-reducing modes
are of relevance, namely the breathing mode $\omega_{b}$, affecting the inter-ion distance, and two
rocking-modes $\omega_{r1}$ and $\omega_{r2}$, occuring perpendicular to the ion distance vector $\boldsymbol{d}$, which describe the shear movement of the ions along two orthogonal axes.
For the investigated two-ion system, the Debye-Waller factor calculates to
$f_{DWF}^{\boldsymbol{r}_1} \approx f_{DWF}^{\boldsymbol{r}_2} = 0.51$, taking
into account the chosen angle between the laser wave vector $\boldsymbol{k}_L$ and
the detector positions $\boldsymbol{r}_i$, $i=1, 2$, as well as the residual phonon numbers
$\left\langle N_{b} \right\rangle \approx 15$, $\left\langle N_{r1} \right\rangle \approx 9$, and $\left\langle N_{r2} \right\rangle \approx 7$ after cooling the two-ion system to the Doppler-limit.
For the cooling laser power employed the saturation parameter is measured to $s = 0.65$ , derived by measuring the Rabi oscillations of the auto-correlation function $g^{(2)}(\tau)$ of a single trapped ion for different laser powers \cite{PhysRevLett.58.203}.
With those values fixed, the only floating parameter is the phase $\delta(\boldsymbol{r}_1)$ of the first recorded photon which is determined by a fit. The corresponding results for $g^{(2)}(\boldsymbol{r}_1,\boldsymbol{r}_2, \tau = 0)$ for each of the four fitted phases $\delta(\boldsymbol{r}_1)$ are shown in Fig.~2(b).
\vspace{0.5cm}

\vspace{1cm}

\end{document}